\documentclass[conference]{IEEEtran}

\usepackage[export]{adjustbox}
\usepackage{booktabs}
\usepackage{float}
\usepackage{flushend}
\usepackage{graphicx}
\usepackage{grffile}
\usepackage{ifpdf}
\usepackage[utf8]{inputenc}
\usepackage{listings}
\usepackage{multirow}
\usepackage[caption=false,font=footnotesize]{subfig}
\usepackage{url}
\usepackage{tabularx}
\usepackage{textcomp}
\usepackage{xcolor}
\usepackage{xspace}
\usepackage{cite}
\usepackage{comment}
\usepackage{outlines}
\usepackage{hyperref}
\usepackage{tikz} 

\definecolor{listinggray}{gray}{0.95}
\definecolor{darkgray}{gray}{0.7}
\definecolor{commentgreen}{rgb}{0, 0.4, 0}
\definecolor{darkblue}{rgb}{0, 0, 0.6}
\definecolor{purple}{rgb}{0.6, 0, 0.6}
\definecolor{middleblue}{rgb}{0, 0, 0.75}
\definecolor{darkred}{rgb}{0.4, 0, 0}
\definecolor{brown}{rgb}{0.5, 0.5, 0}
\definecolor{dkgreen}{rgb}{0,0.5,0}
\definecolor{orange}{rgb}{1,.5,0}
\definecolor{dandelion}{cmyk}{0,0.29,0.84,0}

\usepackage[normalem]{ulem}
\makeatletter
\def\cyanuwave{\bgroup \markoverwith{\lower3.5\p@\hbox{\sixly \textcolor{cyan}{\char58}}}\ULon}
\def\reduwave{\bgroup \markoverwith{\lower3.5\p@\hbox{\sixly \textcolor{red}{\char58}}}\ULon}
\def\blueuwave{\bgroup \markoverwith{\lower3.5\p@\hbox{\sixly \textcolor{blue}{\char58}}}\ULon}
\font\sixly=lasy6 
\makeatother

\def\BibTeX{{\rm B\kern-.05em{\sc i\kern-.025em b}\kern-.08em
    T\kern-.1667em\lower.7ex\hbox{E}\kern-.125emX}}

\newcommand*\circled[1]{\tikz[baseline=(char.base)]{
  \node[shape=circle,draw,inner sep=1pt] (char) {#1};}}

\usepackage{paralist}

\usepackage[acronym]{glossaries}
\glsdisablehyper
\makeglossaries

\newacronym{doe}{DOE}{Department of Energy}
\newacronym{hpc}{HPC}{High Performance Computing}
\newacronym{llnl}{LLNL}{Lawrence Livermore National Laboratory}
\newacronym{ornl}{ORNL}{Oak Ridge National Laboratory}
\newacronym{ml}{ML}{Machine Learning}
\newacronym{ai}{AI}{Artificial Intelligence}
\newacronym{mpi}{MPI}{Message Passing Interface}
\newacronym{rjms}{RJMS}{Resource and Job Management System}
\newacronym{rm}{RM}{Resource Manager}
\newacronym{ecp}{ECP}{Exascale Computing Project}
\newacronym{sdk}{SDK}{Software Development Toolkit}

\definecolor{darkgreen}{rgb}{0,0.5,0}
\definecolor{WildStrawberry}{rgb}{1.0,0.26,0.64}

\newif\ifdraft{}
\drafttrue{}

\ifdraft{}
  \newcommand{\amnote}[1]{ \textcolor{blue} { ***andrem: #1 }}
  \newcommand{\jhanote}[1]{ {\textcolor{red} { ***shantenu: #1 }}}
  \newcommand{\mtnote}[1]{ {\textcolor{orange} { ***matteo: #1 }}}
  \newcommand{\note}[1]{ {\textcolor{red} { ***note: #1 }}}
  \newcommand{\fixme}{{\textcolor{red}{FIXME \xspace}}}
  
  \newcommand{\kyle}[1]{{\textcolor{red}{ Kyle: #1 }}}
  \newcommand{\woz}[1]{{\textcolor{darkgreen}{ [ Woz: #1 ] }}}
  
  \newcommand{\mhnote}[1]{{\textcolor{WildStrawberry}{ [~Mihael: #1~] }}}
\else
  \newcommand{\amnote}[1]{}
  \newcommand{\jhanote}[1]{}
  \newcommand{\mtnote}[1]{}
  \newcommand{\note}[1]{}
  \newcommand{\fixme}[1]{}
  
    \newcommand{\kyle}[1]{}
  \newcommand{\woz}[1]{}
  \newcommand{\mhnote}[1]{}
\fi

\newcommand{\T}[1]{\texttt{#1}\xspace}

\newcommand{\UP}{\vspace*{-1.0em}}
\newcommand{\up}{\vspace*{-0.5em}}

\lstdefinestyle{myListing}{
  frame=single,
  backgroundcolor=\color{listinggray},
  language=C,
  basicstyle=\ttfamily \footnotesize,
  breakautoindent=true,
  breaklines=true
  tabsize=2,
  captionpos=b,
  aboveskip=0em,
  belowskip=-2em,
}

\lstdefinestyle{myPythonListing}{
  frame=single,
  backgroundcolor=\color{listinggray},
  language=Python,
  basicstyle=\ttfamily \footnotesize,
  breakautoindent=true,
  breaklines=true
  tabsize=2,
  captionpos=b,
}

\ifpdf{}
  \DeclareGraphicsExtensions{.pdf, .jpg, .tif}
\else
  \DeclareGraphicsExtensions{.eps, .jpg, .ps}
\fi

\begin{document}
\bstctlcite{IEEEexample:BSTcontrol}

\title{ExaWorks: Workflows for Exascale}


\author{Aymen Al-Saadi$^{1}$, Dong H. Ahn$^{2}$, Yadu Babuji$^{3,4}$, Kyle Chard$^{3,4}$, 
 James Corbett$^{2}$, \\ Mihael Hategan$^{3,4}$, Stephen Herbein$^{2}$, Shantenu Jha$^{5,1}$,  
 Daniel Laney$^{2}$, Andre Merzky$^{5}$,\\ Todd Munson$^{3}$, Michael Salim$^{3}$, Mikhail Titov$^{5}$, 
 Matteo Turilli$^{1,5}$, Justin M. Wozniak$^{3}$ \\
 \small{\emph{$^{1}$ Rutgers, the State University of New Jersey, Piscataway, NJ 08854, USA}}\\
 \small{\emph{$^{2}$ Lawrence Livermore National Laboratory, Livermore, CA 94550, USA}}\\
 \small{\emph{$^{3}$ Argonne National Laboratory, Lemont, IL 60439, USA}}\\
 \small{\emph{$^{4}$ The University of Chicago, Chicago, IL 60637, USA}}\\
 \small{\emph{$^{5}$ Brookhaven National Laboratory, Upton, NY 11973, USA}} \\
}

\maketitle

\pagestyle{plain}

\begin{abstract}


Exascale computers will offer transformative capabilities to combine
data-driven and learning-based approaches with traditional simulation
applications to accelerate scientific discovery and insight. These software
combinations and integrations, however, are difficult to achieve due to
challenges of coordination and deployment of heterogeneous software components
on diverse and massive platforms.
We present the ExaWorks project, which can address many of 
these challenges: ExaWorks is leading a co-design process to create a workflow
\gls{sdk} consisting of a wide range of workflow management tools that can be
composed and interoperate through common interfaces. We describe the initial
set of tools and interfaces supported by the \gls{sdk}, efforts to make them
easier to apply to complex science challenges, and examples of their
application to exemplar cases. Furthermore, we discuss how our project is
working with the workflows community, large computing facilities as well as
HPC platform vendors to sustainably address the requirements of workflows at
the exascale.


\end{abstract}









\section{Introduction}\label{sec:intro}

The coupling of traditional \gls{hpc} with new
simulation, analysis, and data science approaches provides unprecedented
opportunities for discovery but also creates new application and
infrastructure challenges. Several \acrfull{ecp}~\cite{ECP} workflows
exemplify this new reality~\cite{colmena, MuMMI-SC21}: a heterogeneous
combination of applications, \gls{ml} models, and ``glue'' code,
running on heterogeneous compute nodes, orchestrated by a scalable workflow
system. These workflows require specialized workflow management
software which are currently available to only certain large and specialized
inter-disciplinary teams. The gap between capability and requirements will
become more acute with scale and sophistication. Furthermore, ``bespoke''
approaches to workflow development have resulted in many inflexible,
tightly integrated, and stove-piped software solutions. An explosion
in the number of independent solutions is making development and support
of workflows increasingly unwieldy, expensive and unsustainable.


Several technical and non-technical challenges impede the creation of
portable, repeatable, and performant workflows. On the technical side,
workflow management systems (WMS) and complex workflows are difficult to port
and maintain which hinders usability, portability and ultimately adoption. On
the non-technical side, myriad WMS exist which often try to provide complete
and end-to-end capabilities, resulting in duplicated general capabilities but
missing specialized functionality. The lack of coordination, high-quality
specialized and broadly usable common components, has resulted in a disjoint
workflows community that tends towards building ad hoc solutions rather than
adopting and extending existing solutions. The complex workflows landscape
demands an open community-based approach to address these challenges.



The \acrshort{ecp} ExaWorks project was created in response to these
challenges and is addressing both technical and non-technical aspects. We are
co-designing the ExaWorks \gls{sdk} comprised of scalable
workflow tools 
that can be combined to enable diverse teams to produce
scalable and portable workflows for exascale applications. We
do \textit{not} aim to replace the many workflow solutions already deployed
and used by scientists, but rather to provide a well-defined and scalable \gls{sdk}
that provides both common and interoperable APIs, as well as well-tested
workflow technologies, to both the user and workflow communities. 
Most importantly, the SDK will enable sustainability via the creation of a
continuous integration and deployment (CI/CD) infrastructure so that software artifacts produced by participating teams will be easier to
port and maintain.
SDK components will be
usable by other WMSs thus facilitating software convergence in the
workflows community. The ExaWorks SDK is intended to provide scalable
technologies while moving towards sustainability, re-usability, and adoption:

\begin{inparaenum}
    \item \textbf{Re-usability and Composability:} we are partnering with the workflow community to define natural integration points between workflow technologies and begin to define common APIs and reference implementations for capabilities implemented in many workflow systems. 
    
    \item \textbf{Sustainability:} the ExaWorks \gls{sdk} will be included in the Extreme Scale Software Stack (E4S)~\cite{E4S_WWW}, a community effort to provide open-source software packages for developing, deploying and running scientific applications on \gls{hpc} platforms. E4S provides from-source builds and containers of a broad collection of \gls{hpc} software packages.

    \item \textbf{Adoption:} we are building comprehensive \gls{sdk} documentation, with user-facing examples and tutorials, that will facilitate 
    adoption of workflow technologies by developers.

\end{inparaenum}

ExaWorks is also a community-driven project.  
Our vision is to create an open process and community \textit{curated}
\gls{sdk} for workflows: first define interfaces for logical workflow components and
then bootstrap the ExaWorks SDK by adapting a set of existing WMS components
currently being leveraged by \acrshort{ecp} applications. 

In this paper, we present the initial set of ExaWorks technologies and PSI/J, a first component API that aims to provide a unified interface to job schedulers. We then highlight examples of cross-integrations among the current \gls{sdk} technologies and provide highlights from recent application of ExaWorks technologies to extreme-scale workflows.


\section{Understanding HPC Workflows}\label{sec:survey}

Before embarking on the ExaWorks project, we conducted a survey of
ECP application teams to understand their workflow requirements and challenges.  The survey was conducted in two parts: an online questionnaire and targeted deep-dive interviews with a subset of teams. 
In this section we summarize the results and takeaways from this survey~\cite{exaworks-survey}. 


We sent the online questionnaire to 24 \acrshort{ecp} applications teams and the 5 \acrshort{ecp} co-design centers.  We received responses from 15 out of the 29 teams. After reviewing these responses we identified five teams to interview in depth.  Our selection criteria emphasized teams that were developing workflows and that had either written or were leveraging workflow management tools.  Our goal here was to broaden our understanding of these workflows and the tools employed by these teams.

Responses to the survey highlighted that many ECP application teams are orchestrating workflows using homegrown scripts (shell, Python, Perl) and tools like Make. Some teams reported usage of workflow tools:  Airflow, Cheetah, Fireworks, libEnsemble, Merlin, Nexus, Parsl, and Savannah.  Note, we allowed respondents to define ``workflow tool'' broadly, resulting in a mixture of general workflow tools and tools under development for particular sub-domains in HPC.  

We asked teams to describe their workflows. Using these descriptions we grouped responses into the following motifs: 
\begin{enumerate}
\item \textbf{Single simulations:} workflows managing a single simulation, composed of various independent tasks, and often scaling to extreme scale.
\item \textbf{Ensembles:} sets of runs, often statically defined parameter studies, parameter sweeps and convergence studies.
\item \textbf{Analysis:} experiment-driven workflows which involve a mixture of short/small jobs and larger analysis jobs.
\item \textbf{Dynamic:} workflows in which the runs are not known a priori and that involve co-scheduling of disparate tasks and orchestration among tasks. Integrated
HPC and Machine Learning workflows are a growing and important example.
\end{enumerate}

The ensemble motif was the most common motif reported by survey respondents, and often these ensembles were managed via bespoke scripts. 
While one might expect that single simulations would be more common, it is likely that these teams did not employ workflow systems and were thus 
less likely to respond to the survey. 
Analysis and ML/dynamic workflows featured in several responses, and even when general purpose workflow management systems were employed by these teams we found that a significant amount of customized internally developed infrastructure was still required.

We asked respondents to describe the following aspects of their workflows and we summarize the responses here;
\begin{enumerate}
\item \textbf{Internal Orchestration:}
We aimed to understand the need for tasks in a workflow or single batch job allocation to interact with one another. 
Responses indicated use of such coordination, but limited communication
between tasks \textemdash{} though one responding team utilizes
streaming/service oriented workflows where task to task interaction was
required.

\item \textbf{External Orchestration:}
We aimed to understand the extent to which teams utilized multiple machines, or executed workflows across multiple machines.  The responses were evenly divided, with about half of the respondents indicating that their workflows span systems or that they would run them in that mode if they had a workflow tool that makes it possible to do so.  In most cases, the use of multiple systems was driven by the need to scale workloads and to reduce computation time, rather than a differentiation based on hardware or data locality.  Some teams described workloads that exceed scheduler job time limits, requiring submission of several batch jobs, 
and they considered this as a case of external orchestration.

\item \textbf{Homogeneous vs. Heterogeneous tasks:}
In general, most respondents indicated a large dynamic range of job sizes.  Reasons for this range include: scaling/convergence studies, simulation vs. analysis jobs, and co-scheduling of ML and simulation tasks.  Unsurprisingly, we found that teams with more complex and dynamic  workflows reported high levels of task heterogeneity.
\end{enumerate}

The responses and our interviews with teams provided a strong finding that
supporting complex dynamic workflows across multiple machines/data centers,
and porting to new machines is expensive in terms of developer time.  Each
cluster, even those that outwardly appear similar (e.g., Linux OS, Slurm batch
scheduler, etc.), require customization in the workflow.  The subset of ECP
projects that need to run at multiple facilities have developed independent abstraction
layers to support these customizations.   A key takeaway is that attacking the
lower layers of the workflow management stack can bring increased portability
and reduce costs for teams. Finally, a common theme running through
the survey is that developing robust workflows that are fault tolerant and 
portable
is both a pain-point and oftentimes a
determining factor in whether a team will adopt a third party workflow
technology rather than creating their own bespoke capability.  

The aforementioned salient points informed the scope, priorities and approach
of the ExaWorks project (e.g., PSI/J portability layer for schedulers) which
we now discuss.

\section{Three Components of ExaWorks}\label{sec:related}

The three pillars of the ExaWorks technical approach are the robust and performant component technologies, APIs for common workflow components, and the assembly of an open SDK. 
We describe each of these pillars below.

\subsection{Exaworks Components Technologies}


\textbf{Balsam~\cite{balsam2021}} 
provides a hosted platform for orchestrating distributed workflows via web-accessible APIs or Python SDK.  Users define and manage a collection of HPC execution sites that automate wide-area data transfers, resource allocation, and high-throughput, fault-tolerant execution of tasks via pilot jobs.  Since Balsam sites run as user-domain clients communicating with the service over HTTPS, secure deployments are straightforward across diverse platforms.  For instance, Balsam sites can be installed and run on systems ranging from laptops to DOE supercomputers. 
Armed with a collection of sites, clients may then remotely submit tasks to the Balsam API to distribute workloads in near-real time computing scenarios.

The Balsam site's user agent is architected as a collection of platform-agnostic modules (auto-scaling, resource manager synchronization, pilot jobs, data staging) that interface with the underlying systems through a set of adapters implementing \emph{platform interfaces} (Batch Scheduler, MPI Launcher, Compute Resource, Data Transfer Protocol).  Balsam sites can be deployed to new systems by setting the appropriate adapters, while supporting new HPC schedulers or application launch paradigms requires minimal implementation of well-defined interfaces. Pilot jobs dynamically pull workloads from the API and schedule execution of heterogeneous tasks across available resources.  The service maintains a history of state transitions for each task, enabling users to register \emph{pre-/post-processing} or \emph{timeout-/error-handling} hooks, which are invoked by the Balsam site at appropriate stages of the task lifecycle.   


\textbf{Flux~\cite{flux-home}} is a fully hierarchical workload manager for HPC.
It was born out of growing computing needs 
for more sophisticated scheduling and resource management of larger, more heterogeneous and dynamic 
systems at facilities such as DOE national laboratories.
\begin{figure}
  \centering
  \includegraphics[trim=0 0 0 0,clip,width=0.49\textwidth]{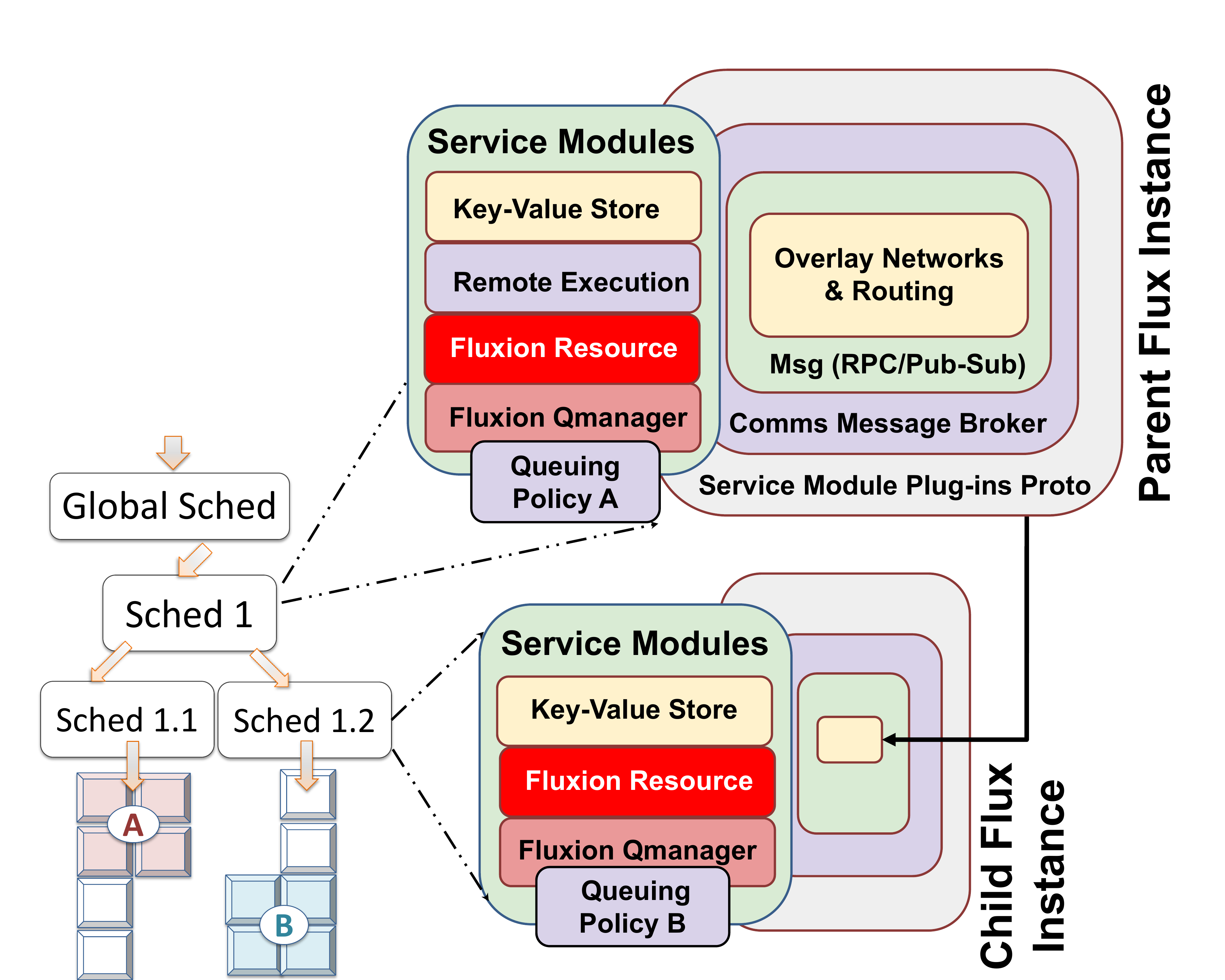}
  \caption{Flux's fully hierarchical software framework architecture}\label{fig:flux}
\end{figure}
Its  fully hierarchical capabilities
have proven to improve scalability and flexibility
significantly through a divide-and-conquer approach
that is well-suited for emerging environments.
Jobs and resources are divided among the Flux instances in the hierarchy
and managed in parallel.

Fig.~\ref{fig:flux}. shows the modular architecture of
Flux, and also depicts how its network can be organized to manage two
Flux instances at different levels of the hierarchy, with a parent Flux instance and
a child Flux instance.
The hierarchical design of Flux provides ample parallelism and flexibility to overcome emerging workflow challenges (e.g., higher job throughput requirement).  Under the
hierarchical design of Flux, any Flux instance can spawn child
instances to aid in scheduling, launching, and managing jobs.  The parent Flux
instance grants a subset of its jobs and resources to each child. This
parent-child relationship can extend to
an arbitrary depth and width, creating many opportunities for parallelization and drastically increasing the
scalability of Flux over traditional schedulers
that rely on a centralized scheme. 
In addition, Flux uniquely provides two modes of operations: single- and multi-user modes.
Emerging scientific workflows often leverage its single-user mode
whereby hierarchical workload management is provided in {\em user space}  within a system-level batch allocation
created by HPC system workload managers, such as
Slurm and IBM LSF.
Such a workload-management overlay allows users to set up their own customized hierarchies
and tune scheduling policies tailored to their workflow.

\textbf{Parsl~\cite{babuji19parsl}} is a parallel programming
library for Python. 
Parsl augments standard Python with workflow constructs
to define dataflow and control semantics. Parsl requires that individual workflow components be implemented as Parsl \textit{Apps}--annotated Python functions that wrap pure Python code or Bash commands. By annotating these functions, Parsl knows that they can be executed concurrently. When apps are invoked, Parsl intercepts the call and returns a \textit{future} in lieu of a result. Developers can call Apps like any other Python functions and link together Apps into sophisticated workflows via standard Python code. Parsl establishes a DAG of dependencies between Apps (based on exchange of data) and sends apps for execution only when dependencies are resolved. 

Parsl implements an extensible runtime model based on 
Python's \textit{concurrent.futures.Executor}~\cite{python-concurrent-futures} interface as a standard
way of executing tasks and a new \textit{provider} abstraction
for managing underlying compute resources. The executor abstraction
has proven to be a flexible interface for integration: Parsl
includes three Parsl-specific executors (HTEX, EXEX, LLEX), a
standard Python ThreadPoolExecutor, and integrations with external
task execution systems such as WorkQueue~\cite{workqueue}, IPyParallel, Balsam, RADICAL-Pilot, Flux, Swift/T, and funcX~\cite{funcx}. Parsl's provider abstraction
offers a Python interface to job schedulers and cloud providers
and currently supports more than one dozen providers, including Slurm, PBS, LSF, AWS, and Kubernetes.

\textbf{RADICAL Cybertools (RCT)} are capabilities developed in Python to
support the execution of heterogeneous workflows and workloads on HPC
infrastructures. RCT provides a workflow engine specialized for the execution
of ensembles (Ensemble Toolkit (EnTK)), a runtime system (RADICAL-Pilot (RP)),
and an interface to batch-systems via RADICAL-SAGA (RS). Here we focus on
RP~\cite{merzky2021design}, which is a pilot system designed to address
research challenges related to efficiency, effectiveness, scalability, and
both workload and resource heterogeneity. RP enables the execution of one or
more workloads comprised of heterogeneous tasks on one or more HPC platforms.
RP offers: (1) concurrent execution of tasks with five types of heterogeneity;
(2) concurrent execution of multiple workloads on a single pilot, across
multiple pilots and across multiple HPC platforms; (3) support of all major
HPC batch systems to acquire and manage computing resources; (4) support of
fifteen methods to launch tasks; and (5) integration with third-party workflow
and runtime systems. The five types of task heterogeneity supported by RP are:
(1) type of task (executable, function or method); (2) parallelism (scalar,
MPI, OpenMP, or multi-process/thread); (3) compute support (CPU and GPU); (4)
size (1 hardware thread to 8000 compute nodes); and duration (zero seconds to
48 hours).

\begin{figure}
  \centering
  \includegraphics[trim=0 0 0 0,clip,width=0.49\textwidth]{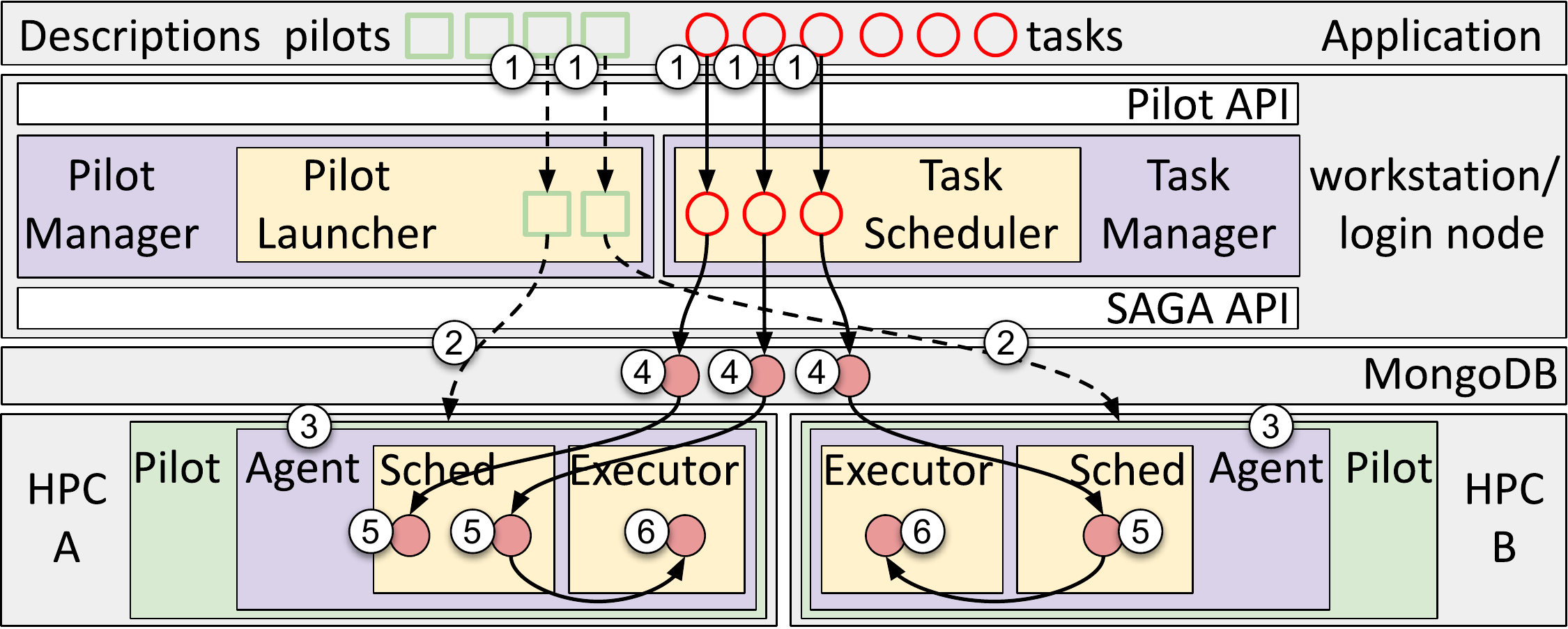}
  \caption{RADICAL-Pilot architecture and execution model}\label{fig:arch-overview}
\end{figure}

RP offers an API to describe both pilots and tasks, alongside classes and methods to manage acquisition of resources, scheduling of tasks on those resources, and the staging of input and output files. Architecturally, RP is a distributed system with four modules: PilotManager, TaskManager, Agent and DB (Fig.~\ref{fig:arch-overview}, purple boxes). Modules can execute locally or remotely, communicating and coordinating over TCP/IP. PilotManager, TaskManager and Agent have multiple components where some are used only in specific deployment scenarios, depending on both workload requirements and resource capabilities. Some components can be instantiated concurrently to enable RP to manage multiple pilots, tasks and HPC resources simultaneously. This allows to scale throughput and enables component-level fault tolerance. Components are coordinated via a dedicated communication mesh implemented with ZeroMQ, which improves overall scalability of the system and lowers component complexity.

\textbf{Swift/T~\cite{Swift_2013}} 
\label{sec:swift}
is an MPI-oriented workflow language and runtime system.  Swift/T is designed to enable the execution of very large numbers of very small tasks across an MPI-enabled computing system.  The tasks could be as simple as short calls to libraries implemented in compiled code, wrapped in scripting languages like Python or R, or packaged as external executables. These tasks can themselves be parallel MPI jobs launched through various mechanisms.  Swift/T has been used to run ensemble applications on the largest available petascale supercomputers, such as COVID-19 population model calibration runs on Theta and deep learning workflows on Summit.

\begin{figure}
\centering
\includegraphics[width=0.45\textwidth,clip]{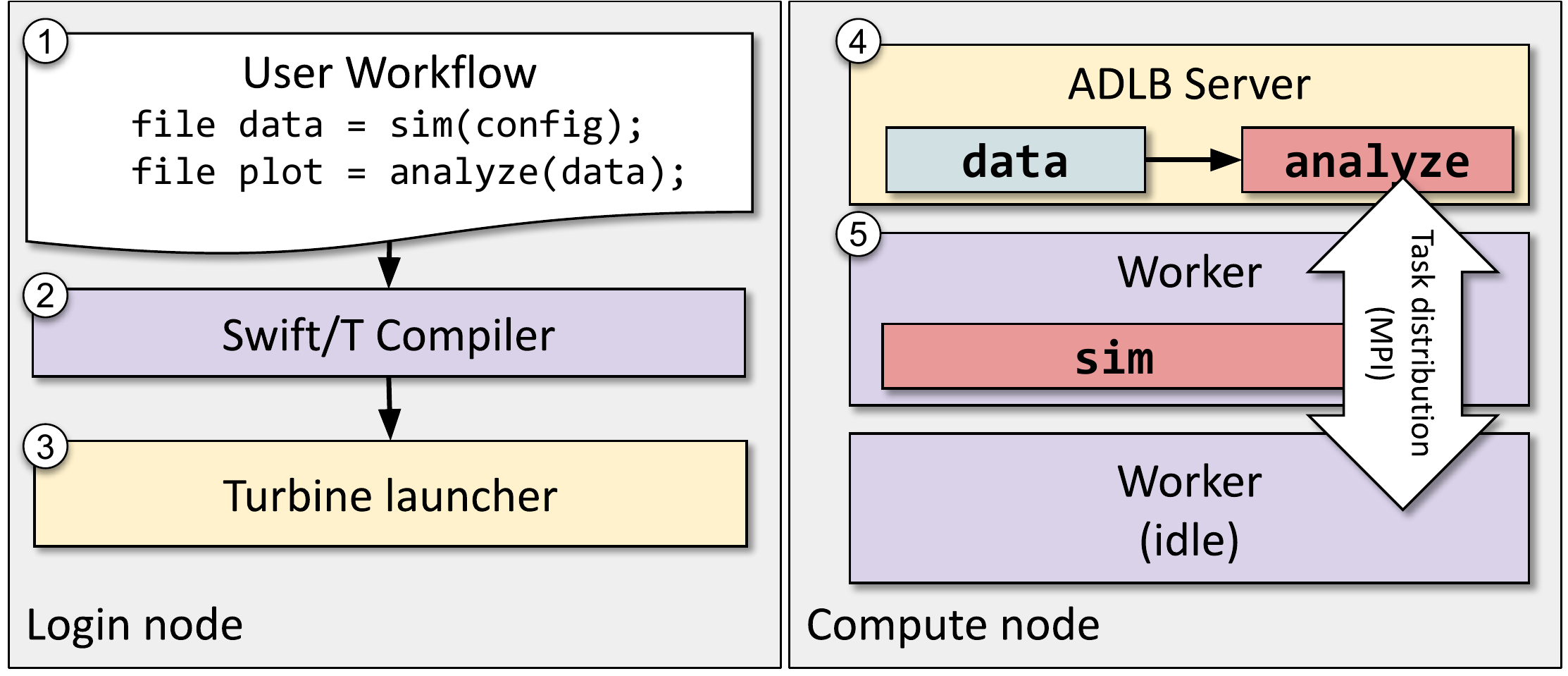}
\caption{Swift/T architecture and data dependency handling \label{fig:arch-swift}}
\end{figure}

Swift/T consists of two components.  Its lower-level Turbine runtime~\cite{Turbine_2013} provides the MPI-level job launch, standard library, configuration features, and optional link-time integration with external scripting languages (Python, R, Tcl, JVM, Julia).  Turbine also wraps around the previously developed ADLB~\cite{ADLB_2010} load balancer, a pure MPI library that performs work-stealing and task distribution.  The higher-level STC compiler is an optimizing compiler that translates the functional Swift language constructs into a format for parallel execution and dataflow-controlled progress.
As shown in \figurename~\ref{fig:arch-swift}, users use the Swift language to develop a workflow program~\circled{1} with implicitly concurrent semantics.  The Swift/T compiler (STC)~\circled{2}~\cite{STC_2014} translates that into a format for execution by the Turbine runtime~\circled{3}, which launches the program in parallel using the MPI implementation, possibly using a system scheduler.  At run time, the ADLB Server~\circled{4} distributes tasks that are ready to run (\T{sim}) to workers~\circled{5}, while tasks that are blocked for data (\T{analyze}) are held until their data dependencies are resolved.

\subsection{Common APIs}

PSI/J is a portability layer across different HPC workload managers 
allowing workflow developers and users to create portable workflows with a standard API. Our survey highlighted the challenge of porting workflows as one of the most critical needs for workflow developers and users alike.  We further observe that most modern HPC workflows each implement their own portability layer, with varying degrees of testing, generality, and performance. 
PSI/J aims to focus effort by pooling into a single layer the collective knowledge obtained from these disparate efforts. PSI/J is composed of both a language-agnostic community-defined specification~\cite{jpsi-spec} and the specification's language-specific implementations.  The high-level goals of the specification are to be lightweight, user-space, minimally prescriptive, scalable via asynchronous operations, general, and extensible to different systems, schedulers, implementations.

The PSI/J specification is broken into three layers that each build on one another.  The first layer forms the base of the specification and focuses on supporting the launching and monitoring of ``local'' jobs (i.e., the client is running on the same system or cluster as the jobs).  The second layer builds on the first to provide support for ``remote'' jobs (i.e., the client must connect over the network, potentially with authentication, to launch and monitor jobs), managing multiple clusters simultaneously, and allowing file staging. The third layer builds on the previous two to support efficient execution of small jobs through the ``nested'', ``pilot job'', or ``jobs-in-jobs'' paradigm. 

As a community, we have focused initially on implementing a PSI/J Python library~\cite{jpsi-python}, but implementations in other languages are encouraged.   The Python library contains the set of
core classes defined by the specification, as well as abstract base classes (ABCs) for components that implement functionality specific to different clusters and batch systems.
The implementation currently supports \texttt{local} for running on a system without a resource manager (e.g., a user's laptop) as well as \texttt{rp} and \texttt{flux} for launching jobs under RADICAL-Pilot and Flux, respectively.  Executor implementations for SAGA~\cite{saga_url} and Slurm~\cite{slurm} are also under development.

\subsection{Exaworks Software Development Kit~\cite{Exaworks_SDK_GH}}


The ExaWorks ~\gls{sdk} aims to make 
workflow technologies easier to deploy, build upon in diverse applications, leverage multiple workflow systems for the same application, and interoperate with external systems.
This will democratize access to increasingly hardened, scalable, and portable workflow technologies, components, and solutions to typical problems.

The SDK is implemented via a community-based approach. We follow an open community-based design process in which all artifacts are tracked using an open process on GitHub. We have defined community policies for inclusion of technologies in the SDK~\cite{sdk-policies}  to ensure minimum standard software quality practices for reliable deployment and use, modeled on E4S. 
We will work with workflows developers to integrate technologies that meet these policies.
For example, we require that technologies have Docker containers and Spack packages for deployment. 
These packaging solutions allow for portable and reproducible deployment and testing techniques, which vary considerably among workflow tools.  
We have defined a Fork+Pull model-based GitHub development workflow and and GitHub Actions-based continuous integration testing and deployment for SDK components.

Importantly, we aim for the SDK to facilitate progressively advanced levels of interoperation among the tools. 
We define three interoperability levels as follows:

\begin{itemize}
\item \textbf{Level 0: Technologies can be packaged together:}  A basic container or other deployment system can support technologies in the same environment.  
\item \textbf{Level 1: Technologies can interoperate:}  A single workflow solution can use features from two or more systems which use tool-specific interfaces.
\item \textbf{Level 2: Sustainable interoperability:}  Users can perform deep customization of workflow system behavior, choosing from and composing tools that interoperate through the common APIs.
\end{itemize}


The next steps for the SDK are to develop and harden deeper integration of our technologies, to extend our CI/CD pipeline to \gls{ecp} systems,
and to integrate othre community workflows tools.
We expect that a richer CI/CD pipeline and more examples of deeper integration examples will significantly facilitate the rapid integration of a wider range of community workflows libraries and tools into the \gls{sdk}.
Thus far, we have focused on small-scale interoperation, but we plan to extend our solutions to ensure scalable integration of the components on pre-exascale and early access exascale systems for immediate readiness as exascale systems become available. We also plan to apply our \gls{sdk} solutions to \gls{ecp} and other exascale-relevant workflows including the \gls{ecp} ExaAM workflow.

\section{Integration Examples}\label{sec:examples}


The ultimate goal of the ExaWorks SDK is to enable various components to be adopted and combined to meet use cases. Towards this goal we have prototyped integrations between  
SDK tools to validate the feasibility of this approach and to enhance these tools with new capabilities. 

\begin{figure*}[t]
	\up
	\centering
	\subfloat[][]{
	  \includegraphics[width=0.30\textwidth]{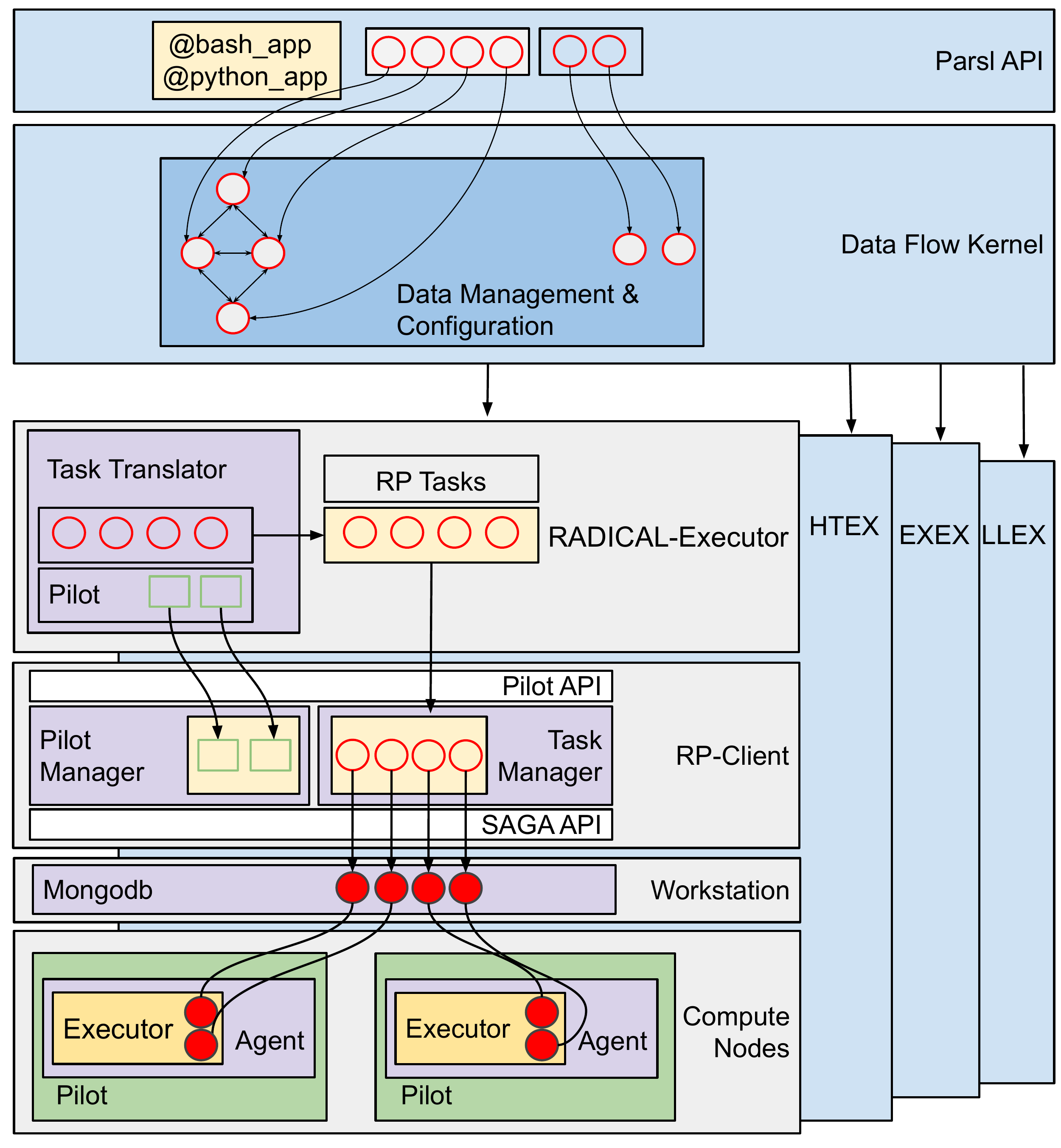}
	  \label{fig:rp-parsl-int-arch}}
		\hfill
	\subfloat[][]{
	  \includegraphics[width=0.30\textwidth]{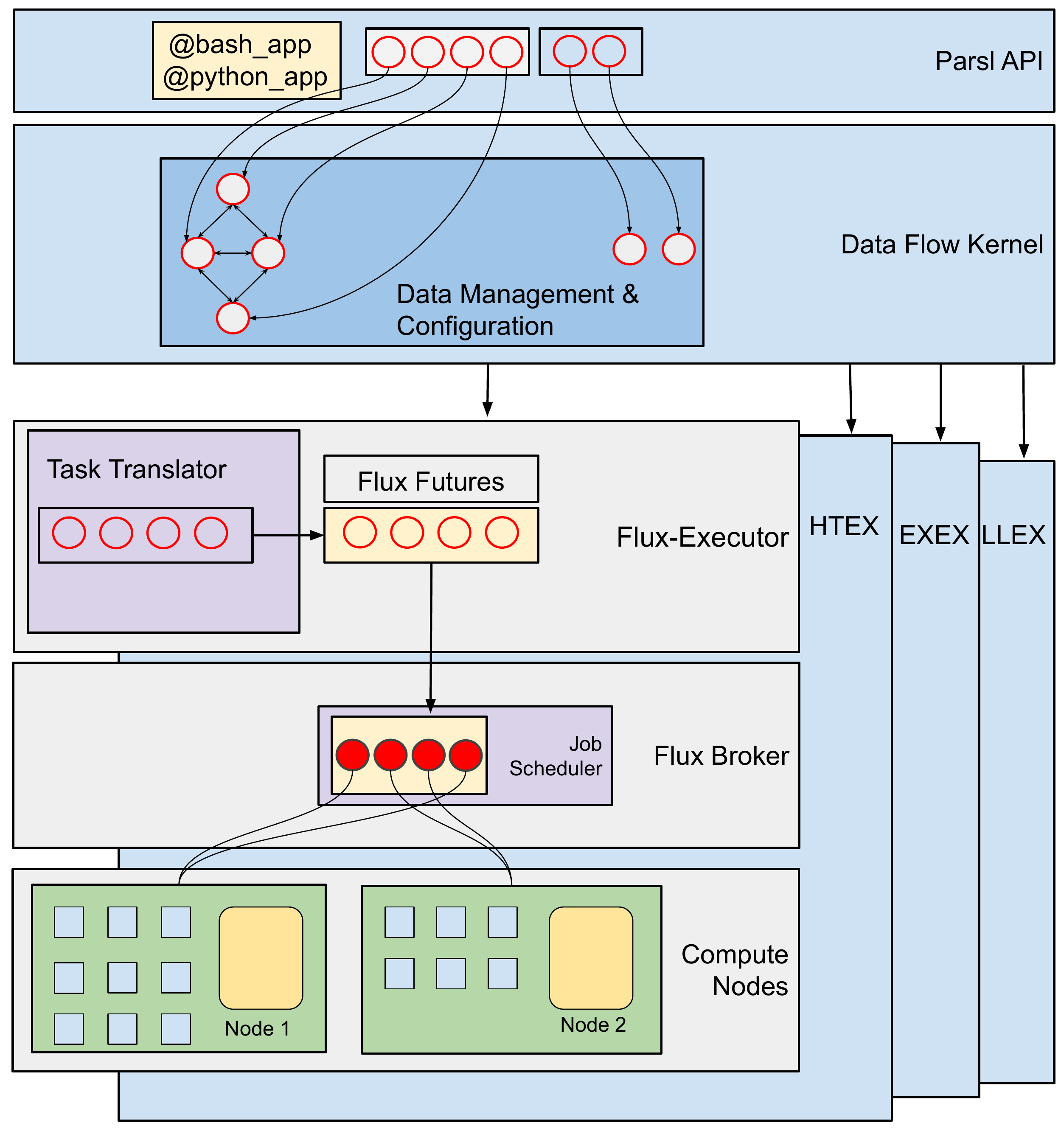}
	  \label{fig:flux-parsl-int-arch}}
	  	\hfill
	\subfloat[][]{
	  \includegraphics[width=0.30\textwidth]{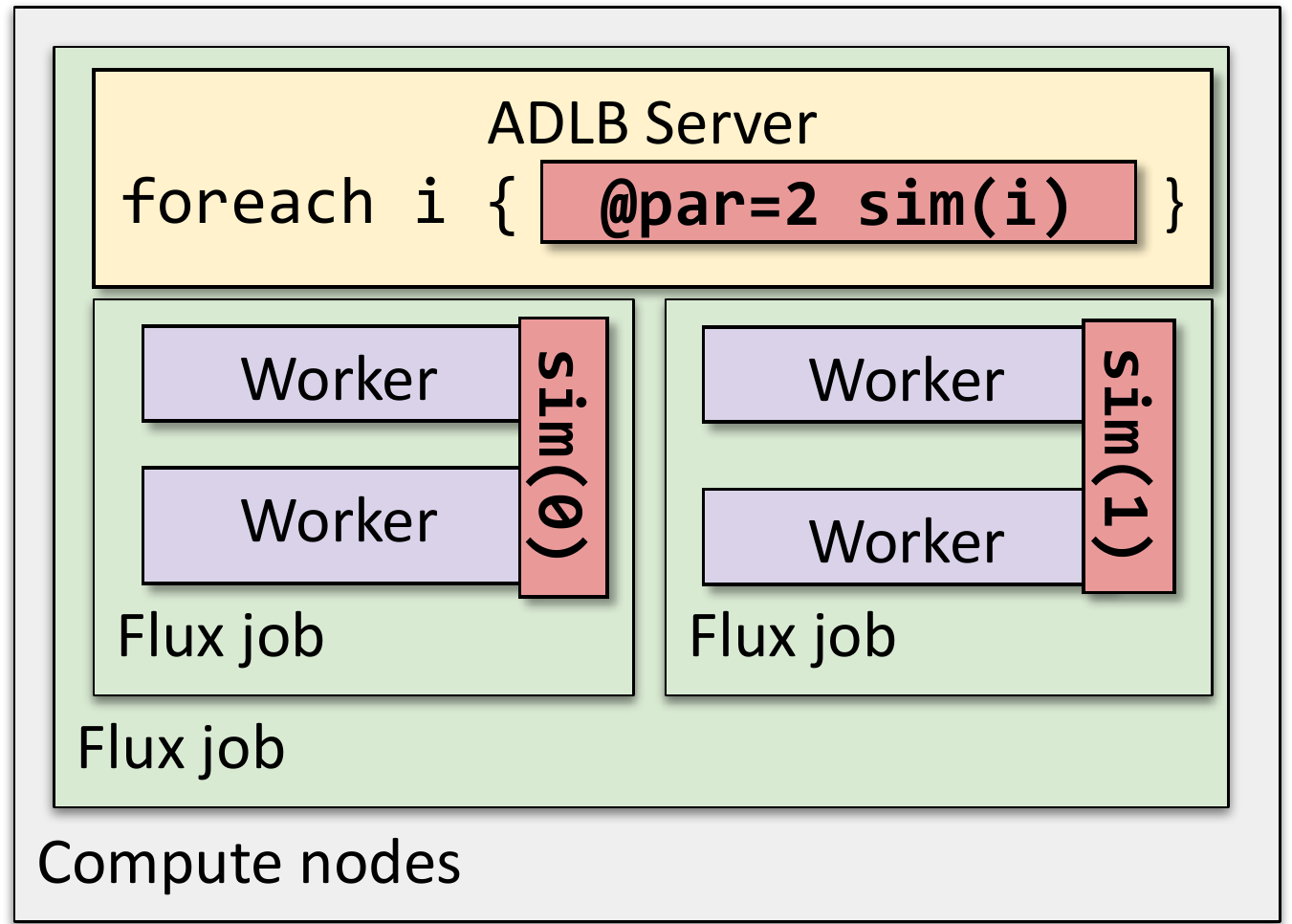}
	  \label{fig:swift_flux}}

	  \caption{Integration architectures: (a) Parsl + RADICAL, (b) Parsl + Flux and (c) Swift/T  +  Flux }
	\label{fig:integration_archs}
	\UP
  \end{figure*}

\textbf{Parsl + Flux:}
Parsl's standard executors are not designed to schedule tasks based on resource requirements. To provide this capability we integrated Parsl and Flux such that Parsl can leverage Flux's scheduling features and support applications with varying resource requirements. 
To do so, we implemented Parsl's standard executor interface in Flux as shown in \figurename~\ref{fig:flux-parsl-int-arch}.

\textbf{Flux + RADICAL-Pilot:}
We extended RADICAL-Pilot (RP) to support Flux as an alternative backend
system to place, launch, and manage tasks across allocated resources. The
delegation of these costly operations to Flux helps to reduce RP runtime
overheads. RP can increase the overall task throughput by launching multiple
Flux instances within the same job allocation and using them
concurrently\cite{lee2020scalable}. RP starts with bootstrapping its
components in the job allocation, then it launches Flux instances and
schedules tasks on them for execution. Flux schedules, places and launches
tasks on compute nodes via its daemons. RP Executor tracks task completion,
and communicates this information to RP Scheduler, based upon which RP
Scheduler passes more tasks to Flux for execution.

\textbf{Parsl + RADICAL-Pilot:}
We integrated RP and Parsl, as shown in \figurename~\ref{fig:rp-parsl-int-arch}, to provide a new scalable runtime in Parsl that is capable of managing and executing heterogeneous tasks efficiently. 
We designed this integration based on the structural adapter pattern~\cite{Ayeva_mastering_2018}. The adapter pattern allows Parsl and RP 
to communicate seamlessly using efficient object conversion at execution time. 
Our performance characterization of RP-Parsl~\cite{alsaadi_RADICAL_2021} showed that the overheads of RP-Parsl are small and invariant to the number of tasks, and number/type of resources.


\textbf{Swift/T + Flux:}
We have designed an integration between Swift/T and Flux.
Previously, 
we have developed techniques for managing large numbers of small to medium-sized MPI jobs: \T{Comm\_create\_group()}~\cite{Swift_MPI_2013} and \T{Comm\_launch()}~\cite{Launch_2019}, and we recognize that these features could be extended by tapping into Flux's abilities to handle a workload consisting of hierarchical or nested parallelism.
For example, Swift/T could be used to specify the outermost parallel program, and parallel simulations could run inside it.  More complex structures are also possible.
This model is depicted in \figurename~\ref{fig:swift_flux} where Swift/T runs inside a 
Flux instance.


\section{Use Cases and Applications}\label{sec:applications}

We outline exemplar use of ExaWorks technologies in ECP applications and 
in extreme-scale COVID-19 research.

\subsection{Representative Applications}

We are working with several ECP application teams to understand requirements 
and apply ExaWorks technologies.

\textbf{Cancer Distributed Learning Environment (CANDLE)}
is a deep learning (DL)-oriented cancer application suite for exascale.  It consists of three key software products: Benchmarks, Libraries, and Supervisor~\cite{CANDLE_2018}.  The Benchmarks are a collection of relatively small, self-contained cancer applications, such as predicting the presence of a tumor in an RNA expression sample.  The Libraries are a collection of tools to support DL, including configuration, I/O, checkpoint/restart, and analysis methods.  Supervisor is the workflow component of CANDLE and is built on Swift/T.  Supervisor workflows include flat bag-of-tasks cases, hyperparameter optimization~(HPO), and more advanced data-oriented investigations~\cite{CANDLE_2020}.  The workflows allow the Benchmarks and Library features to be easily plugged in and run on a wide range of pre-exascale resources.
For example, a new HPO idea could easily be applied to CANDLE-compliant Benchmarks for evaluation, and new CANDLE-compliant Benchmarks (or other DL models) can easily be developed and run at scale.

\textbf{Colmena} is an open-source Python framework for ML-steering of simulation campaigns at scale. Colmena allows developers to define multi-fidelity
simulations---either computational or ML surrogates---for determining properties.  Colmena further allows users to define the control logic
used to select which which simulation and ML tasks to execute when, as well as implementations of those tasks.  Colmena allows for different steering functions to be defined and used to orchestrate the campaign, typically these functions are themselves ML algorithms that are repeatedly trained and applied. Colmena manages the complexity of task dispatch, results collation, ML model invocation, and ML model (re)training, using Parsl to execute tasks on HPC systems.  Colmena has been used to drive electroltye design campaigns spanning an enormous design space and using thousands of nodes on the Theta supercomputer. The ML-guided steering approach is able to accelerate discovery of molecules by several orders of magnitude when compared with unguided searches.

\textbf{Exascale Additive Manufacturing (ExaAM)} 
is building a complex workflow to simulate a laser melt-pool additive manufacturing build process. The workflow is composed of several application codes, each capturing a different scale or physics of the problem.  The initial ExaAM challenge problem is comprised of five phases: a full-build continuum model to set initial conditions, a hi-fidelity melt pool model that captures powder bed fusion, a detailed microstructure-scale solidification model, a polycrystal property model, and finally a continuum model that leverages the AM process-aware material model from the detailed fine-scale simulations.  Each stage feeds information to the next stage, and iteration can occur across stages.  Some stages leverage CPU's and others GPU's. Several stages are themselves small workflows typically leveraging the \textit{ensemble} motif. A nuance of the model is that it follows the additive process, which means that it is possible to parallelize computations over space and time by dividing the build process into layers and then breaking the laser path into segments for each layer.  Thus, the ExaAM workflow requires integration of multiple physics applications and when executed at full scale will require complex orchestration of many sub-workflows.  Initial integration of ExaWorks SDK technologies has shown that improvements in throughput are possible with minor changes to existing scripts.  The ExaAM and ExaWorks teams have identified additional stages in the workflow that could benefit by incrementally integrating ExaWorks SDK capabilities to improve portability and performance.

\subsection{Gordon Bell}

The winner and two of three finalists of the SC20 Gordon Bell Special Award
for COVID-19 competition leveraged ExaWorks technologies. 
We believe that this is a demonstration of the effectiveness of 
the use of high-quality scalable workflow building blocks to create sophisticated dynamic workflows that can leverage leadership-class supercomputers. 
All four COVID-19 award finalists involved workflows, and three of them 
used ExaWorks technologies. 
Each team developed their own tailored workflow solutions, leveraging ExaWorks technologies at key points, to enable scalability while reducing the developer time needed to support the scale and magnitude of these research efforts.”

For example, the winner of the award addressed the challenge of evaluating a
potentially huge set of ``biologically interesting” conformational changes by
creating ``a generalizable AI-driven workflow that leverages heterogeneous HPC
resources to explore the time-dependent dynamics of molecular systems.” This
workflow used DeepDriveMD and components from the ExaWorks SDK. It combined
cutting-edge AI techniques with the highly scalable NAMD code to produce a new
high watermark for classical MD simulation of viruses by simulating 305
million atoms. The ORNL Summit system was able to deliver impressive sustained
NAMD simulation performance, parallel speedup, and scaling efficiency for the
full SARS-CoV-2 virion. AI helped identify interesting conformational changes
that were explored further in detail to understand the important molecular
changes that occur due to the ``jiggling and wiggling of atoms”.

Flux was used by another finalist doing drug design to provide the scalable backbone of Livermore’s Rapid COVID-19 Small Molecule Drug Design workflow. 
The use of Flux provided an unprecedented level of composability of workflow systems in such a way that highly complex campaigns such as drug design are easily architected in a timely fashion.

The third finalist adopted Swift/T to develop a highly scalable
epidemiological simulation and machine learning (ML) platform. The workflow
was a complex structure of CityCOVID, a parallel RepastHPC agent-based
modeling simulation of the 2.7 million residents of Chicago, interspersed
with batteries of ML-accelerated optimization tasks.
Integrating the complete CityCOVID and ML epidemiological modeling platform was aided by multiple Swift/T design features. The simulation itself is a stand-alone C++ module that, in this case, ran on 256 cores and communicated internally with MPI. Invoking large, concurrent batches of such runs efficiently is one of the main capabilities of the Swift/T runtime, which invoked the simulator repeatedly through library interfaces.


Another challenge was generating and coordinating the large number of single-node optimization tasks, each 
used vendor-optimized multithreaded math kernels. These single-node tasks were calls to a range of R libraries, dispatched via a custom R parallel backend. This extended the notion of workflow composability, a key theme of the ExaWorks project, into the algorithmic control of the simulation and learning through external algorithms developed in “native” ML languages R and Python. This was implemented using the resident, or stateful, task capabilities of Swift/T and the associated EMEWS algorithm control framework.



\section{Building a Workflows Community}\label{sec:community}


In collaboration with the NSF-funded WorkflowsRI project, we are hosting
a series of workflows community summits that aim to bring the diverse
workflows community together. 

The first summit~\cite{summit_1} brought together 48 international participants representing many WMSs, with the goal to identify crucial challenges in the workflows community. The summit considered six broad themes: FAIR workflows, training and education, AI workflows, exascale challenges, APIs and interoperability, and developing workflow community. 

The second summit~\cite{summit_2} focused on technical approaches for realizing many of the challenges identified in the first summit. It included 75 workflows developers 
and focused on three core topics: 
defining common workflow patterns and benchmarks, identifying paths toward interoperability of workflow systems, and improving workflow systems’ interface with legacy and emerging HPC software and hardware.





\section{A Vision for the Future}\label{sec:conclusion}



Workflow system software will become necessary components of HPC software
stacks. Including workflow requirements in the procurement/requirements
process for future HPC platforms and Cloud environments will aid efforts to
establish community standardization around workflow APIs. This will require
user and facility community engagement around defining  common APIs, and
eventually, widely used and shared implementations. Encouraging adoption of
these common APIs in the user community while advocating for workflow
requirements to enter directly into the procurement processes at facilities,
will lay the foundation  for workflow developers to build, maintain, and
support their workflow technologies in partnership with the facilities and
private sector. This will enhance the sustainability of workflow system
software. Of course, for this adoption and partnering to occur, common
workflow APIs and reference implementations must be widely ported and
extensively tested to meet user expectations.


PSI/J represents an initial effort towards this goal, in that it is scoped to
achieve both adoption by bespoke workflow developers (i.e., small teams of
domain scientists), inclusion in workflow tools (starting with the
ExaWorks SDK), and be tested  widely across many facilities and cloud
providers. The focused scope of PSI/J may allow for it to be included as a
requirement in future procurements, with the possibility of it becoming a
standard API provided for both users and workflow system developers.


Partnering with industry, including cloud vendors, is a key aspect of a plan
to sustain a workflows SDK beyond its current \acrshort{ecp} funding.  Cloud vendors are increasingly providing workflow capabilities as
well as HPC capabilities. The commercial ModSim community involved in engineering and product development, are rapidly moving to
leverage cloud capabilities in their software platforms.  It is therefore
important that the HPC community engage with both HPC system integrators, as
well as cloud vendors, to socialize and propagate performant workflow
technologies. 


We will release a first version of the SDK in 2021 and then periodically release new versions as capabilities are added.  Subsequent releases will include increasing use of \acrshort{ecp} continuous integration capabilities and deployments on additional facilities.  PSI/J development and releases will continue as more backends are added and it is deployed and tested at data centers.  We will be co-organizing community workshops and invite the wider community to participate. Finally, all ExaWorks development activities will continue to be hosted on GitHub and are open for community participation. 

We have made concrete progress at creating an SDK of workflow technologies
focused on exascale workflow requirements.  Our focus on establishing a
rigorous continuous integration and deployment workflow, as well as our
engagement with DOE facilities, is a key part of our vision to create a
community curated workflows SDK. Our vision is lofty; our mandate is lucent:
instantiate and grow ExaWorks into a community-owned and guided body that can
contribute \texttt{de facto} standards for HPC and Cloud workflows, host
reference implementations of common APIs, and curate an SDK of industrial
strength components focused on running scalably on the most powerful HPC
platforms available.  Succeeding at this vision will depend as much on
technical capabilities as on engagement with stakeholders. We invite the
workflows community to participate and collaborate with us as we work to make
this vision a reality.

\footnotesize
\noindent {\bf Acknowledgements} {
This research was supported by the Exascale Computing Project (17-SC-20-SC), a
collaborative effort of the U.S. Department of Energy Office of Science and
the National Nuclear Security Administration. This work was performed under
the auspices of the U.S. Department of Energy by Lawrence Livermore National
Laboratory under Contract DE-AC52-07NA27344 (LLNL-CONF-826133), Argonne Natonal Laboratory under Contract DE-AC02-06CH11357, and Brookhaven National Laboratory under Contract DESC0012704. We acknowledge
the guidance of the ExaWorks Advisory Committee:  Debbie Bard (LBNL/NERSC),
Ian Foster (ANL/Chicago), Jack Wells (NVIDIA), Mallikarjun Shankar (OLCF/ORNL),
Bill Allcock (ALCF), Daniel S. Katz (UIUC), and former members Robert Clay
(formerly of SNL) and Justin Whitt (ORNL). We thank Rafael Ferreira da Silva
and Henri Casanova of the Workflows-RI project for partnering to organize community summits. 

\bibliographystyle{IEEEtran}
\bibliography{main, exaworks}

\end{document}